\documentclass[prd,showpacs,preprintnumbers,floatfix,superscriptaddress,10pt]{revtex4}
\usepackage[mathscr]{eucal}
\usepackage[dvips]{color}
\usepackage[dvips]{graphicx}
\usepackage{epsf}
\usepackage{bm}
\usepackage{amssymb}
\usepackage{amsmath}
\usepackage[normalem]{ulem}
\newcommand{\be}{\begin{equation}}
\newcommand{\ee}{\end{equation}}
\newcommand{\ba}{\begin{eqnarray}}
\newcommand{\ea}{\end{eqnarray}}
\newcommand{\ban}{\begin{eqnarray*}}
\newcommand{\ean}{\end{eqnarray*}}

\newcommand{\ie}{{\it i.e.\,}}

\newcommand{\bef}{\begin{figure}}
\newcommand{\eef}{\end{figure}}
\newcommand{\bce}{\begin{center}}
\newcommand{\ece}{\end{center}}

\begin{document}

\title{Chromohydrodynamical instabilities induced by relativistic jets
}
\author{Massimo~Mannarelli}
\author{Cristina~Manuel}
\affiliation{Instituto de Ciencias del Espacio (IEEC/CSIC),
Campus Universitat Aut\`onoma de Barcelona, Facultat de Ci\`encies, Torre C5 E-08193 Bellaterra (Barcelona), Spain}
\date{\today}
\begin{abstract}
We study the properties of the chromohydrodynamical instabilities
induced by a relativistic jet that crosses the quark-gluon plasma.
Assuming that the jet of particles and the plasma
can be described using a hydrodynamical approach,
we derive and discuss the dispersion laws for the unstable collective modes.   In our analysis  the chromohydrodynamical equations for the collective modes   are tackled in the linear response approximation. Such an approximation,  valid for short time scales, allows   to study in a straightforward way the dependence of the dispersion laws of the collective modes on the velocity of the jet, on the magnitude of the momentum of the collective mode  and on the angle between these two quantities. In the conformal limit  we find  that unstable modes arise for  velocity of the jet  larger than the speed of the sound of the  plasma and only modes with momenta smaller than a certain values are unstable.  Moreover, for
ultrarelativistic velocities of the jet  the longitudinal mode becomes stable and the most unstable modes correspond to relative angles between the velocity of the jet and  momentum of the collective mode larger than $\sim \pi/8$.
Our results suggest  an alternative mechanism for the description of the jet quenching  phenomenon, where the jet crossing the plasma loses energy exciting colored unstable modes.
\end{abstract}
\preprint{}
 \maketitle

\section{Introduction}
One of the remarkable findings at the Relativistic Heavy Ion Collider, RHIC, is that hydrodynamical models
are able to  accurately describe  the spectra of soft  hadrons
produced in  heavy-ion collisions \cite{Adams:2005dq,Kolb:2003dz}. 
Hydrodynamics, being an effective macroscopic
approach valid at large distances and long time scales, cannot give account of the pre-equilibrium
stage of the reaction. However,  hydrodynamics describes correctly
elliptic flow at RHIC, and  allows  to put some bounds on the thermalization time $t < 1$ fm/c.

An  experimental evidence of the production  of
a thermalized quark-gluon plasma  in ultrarelativistic heavy-ion
collisions at RHIC \cite{Adams:2005dq}  is the so-called
jet quenching. In this phenomenon  high $p_T$ partons produced in the initial stage of the collision by hard scatterings   loose
energy  and degrade, mainly by radiative processes, while traveling through the hot and dense medium transferring energy and momentum  to the plasma (see \cite{Kovner:2003zj} for reviews).
 In order to describe these processes and the subsequent modification of the hadronic spectra due to the interaction of the high $p_T$ partons with the medium, various models have been proposed where perturbative QCD is supplemented with medium-induced parton energy loss \cite{Eskola:2004cr} or where the AdS/CFT correspondence \cite{Liu:2006ug} is employed.

It is interesting to analyze whether some aspects of the process of jet quenching can be described employing a hydrodynamical picture \cite{Chaudhuri:2005vc}. Hydrodynamics describes  the behavior of a system in local equilibrium in terms of conservation laws of macroscopic quantities. On the other hand, the jet quenching is related to the modification of high-$p_T$ parton fragmentation processes in the medium. It is then clear that hydrodynamics cannot describe the microscopic processes related to jet quenching, but it can give information on the macroscopic and collective behavior of the system composed by the  plasma and the hard jets.

In Refs.~\cite{Stoecker:2004qu,Casalderrey-Solana:2004qm}  the effect of a  jet crossing the plasma at high speed  has  been analyzed. The authors have employed a hydrodynamical approach to describe the process of   energy and momentum
deposition from  the fast jet to the surrounding  plasma.
In  a hydrodynamical picture, a high $p_T$ jet crossing the medium at a velocity higher than the speed of sound  forms shock waves with a Mach cone structure. Such shock waves should be detectable in the low $p_T$ parton distributions
 at angles $\pi \pm 1.2 $ with respect to the direction of the trigger particle.
 A preliminary analysis of the azimuthal dihadron correlation performed by the STAR Collaboration \cite{Adams:2005ph}
 and the PHENIX Collaboration \cite{Adler:2005ee} seems to suggest  the formation of such a conical flow. However,  it is yet unclear whether
those structures are compatible with Mach cone formation and alternative explanations  have been proposed \cite{Polosa:2006hb}.
The study performed in Ref. \cite{Chaudhuri:2005vc} concludes that for realistic phenomenological values
of the hydrodynamical variables the Mach cone effects are too weak to explain the  PHENIX results.
It is our aim here to explore  whether other hydrodynamical effects, not
considered in \cite{Chaudhuri:2005vc}, may enhance the  signal.

The study of the interaction of a relativistic stream of particles with an electromagnetic plasma is a topic of interest in different fields of physics, ranging from inertial confinement fusion,
astrophysics and cosmology. When the particles of the stream are charged, plasma instabilities develop,
leading to an initial stage of fast growth of the electromagnetic fields. One then talks about
filamentation, two-stream or Weibel instabilities, according to which is the fastest growing collective mode, although the notation is non-universal and sometimes confusing. Weibel instabilities usually refer to the case
when the plasma is out of equilibrium, and with an anisotropic distribution in the velocities of its
constituents, but the name is also  used in the context of a jet moving in an equilibrated plasma
if the instability appears in the transverse modes. These sort of jet induced instabilities have been studied
using a variety of methods, from kinetic theory to hydrodynamics \cite{Honda,Bret}. Experimental evidence of the relativistic
filamentation instability has also been reported in Ref.~\cite{exp-inst}.

The study of chromo-Weibel instabilities is now  a very active field of research 
\cite{Mrowczynski:1994xv}-\cite{Bodeker:2007fw}
(see also the recent reviews
\cite{Mrowczynski:2006ad} and \cite{Strickland:2007fm} for a more complete list of references).
This is so because it was suggested \cite{Mrowczynski:1994xv,Arnold:2004ti} that the presence of plasma instabilities could be a natural explanation for the
fast equilibration that measurements of the elliptic flow at RHIC seem to imply. In the early stage of
a heavy-ion collision, the non-equilibrium anisotropic distributions of the partons should be responsible
for the fast growth of the chromomagnetic plasma modes, which in turn would isotropize the system and speed
up the thermalization process. Whether the experimental conditions met at  RHIC are favoring this thermalization scenario or not is a question that requires hard numerical simulations.

In this paper we study how a relativistic jet crossing an equilibrated quark-gluon plasma induces instabilities using a chromohydrodynamical approach \cite{Manuel:2006hg}. We will assume that both the plasma and the jet  can be described using hydrodynamics. Therefore, regarding the plasma, we  consider conditions that can be  realized  
after $t \sim 1$ fm/c of the moment the heavy ion collision has taken place and the plasma has thermalized.  Regarding the jet, there are some experimental evidences that the particles that constitute the jet that crosses the plasma equilibrate. Indeed, consider a fast parton that traverses the medium in the the direction opposite to the direction of a high $p_T$ trigger particle. The mean value of the momentum associated of the soft hadrons emitted by the parton that crosses the medium and detected in the hemisphere opposite to the direction of the trigger particle    reaches a common value \cite{Adams:2005ph} suggesting that the energy lost by the fast parton  thermalizes \cite{Chaudhuri:2005vc}. 

In studying the evolution of the system composed by the plasma and the jet we will employ  ideal hydrodynamic equations that do not take into account the effect of collisions. Therefore our results will be valid on time scales shorter than the mean free path time for collisions. We postpone the study  of 
more realistic fluid equations to future work.

The validity of the chromohydrodynamical approach will be our starting  assumption, as this allows
us  to simplify the equations governing the evolution of the system. To the best
of our knowledge, only Ref.~\cite{Pavlenko:1991ih} considers the
possibility of the appearance of filamentation instabilities
produced by hard jets in heavy-ion collisions. However, the approach considered there and
ours  are  different. The analysis of
Ref.~\cite{Pavlenko:1991ih} is performed within kinetic theory,
and thus relies on the quasi-particle picture and a weak coupling
scenario. Instead, we use fluid equations. We note that  colored plasma
waves are believed to be very quickly damped (and this is the
reason to exclude them in the study of
Ref.~\cite{Chaudhuri:2005vc}). That conclusion might have to be
reviewed, as in the presence of instabilities rather than damping,
one finds exponential growth of the fields. Our approach is also
different to the color wake field scenario of
Refs.~\cite{Ruppert:2005uz,Chakraborty:2006md}. In those references, the authors studied the field response to a single moving
colored particle, which produces a color charge density wake.
Instead, we consider a color neutral jet, treated hydrodynamically, that
moves trough the plasma and  produces color fluctuations and
instabilities.

This paper is structured as follows. In Section \ref{sec-eqs} we review the chromohydrodynamical equations describing
the fluctuations of a plasma around the stationary colorless state in the  linear approximation
\cite{Manuel:2006hg}. In Section \ref{sound-sec}  the equation describing sound colorless fluctuations and plasma colored fluctuations are derived. In Section \ref{plasma+jet-sec} we consider the case where  a relativistic jet  crosses the plasma. Assuming that both the plasma and the jet can be treated using a hydrodynamical approach, we derive the dispersion law for the collective modes. We find that
there is one unstable  mode if the velocity of the jet is larger than the speed of sound and if the momentum of
the collective mode is in modulus smaller than a threshold value.
Regarding the orientation of the momentum of the collective mode, we study separately the case where it is  collinear with the velocity of the jet, the case
where it is orthogonal and then for a generic angle between the two. Quite interestingly we find that  the unstable modes with momentum parallel to the velocity of the jet is the dominant one for velocity of the jet $v \lesssim 0.8$.
For larger values of the jet velocity only the modes with angles larger than $\sim \pi/8$ are significant and the dominant unstable modes correspond to  angles $\sim \pi/4$. We draw our conclusions in  Section \ref{conclusion-sec}. We work using natural units $\hbar = c = k_B = 1$ and use metric convention $(1,-1,-1,-1)$.

\section{Chromohydrodynamic equations for the quark-gluon plasma}
\label{sec-eqs}

Hydrodynamical equations are the
expressions of the conservation laws of a system when it is in
local equilibrium. In the quark-gluon plasma there are conservation
laws which concern the baryon current, the color current and the
energy-momentum tensor.
 In Ref.~\cite{Manuel:2003zr} the local
equilibrium state for the quark-gluon plasma has been determined.
It is in general described by one singlet four velocity, a baryon
density, singlet energy and pressure, and in principle, a
non-vanishing color density. However, dynamical processes
associated to the existence of Ohmic currents tend to whiten the
plasma quickly, on time scales much shorter than  momentum
equilibration processes \cite{Manuel:2004gk}. For this  reason one
can expect that only  colorless (singlet) fluctuations are relevant at large time and space scales.
However, there are situations,  as the one considered in the present paper, when color fluctuations grow on short time scales instead of being damped. Therefore in order to describe  the short time evolution of the plasma, one needs to include color hydrodynamical fluctuations in the equations.  
In  Ref.~\cite{Manuel:2006hg} such a  chromohydrodynamical approach  for  the short time evolution
of the system has been formulated,  and here we will review the basic set of equations.

For simplicity, as in Ref.~\cite{Manuel:2006hg}, we will only consider
the contribution of quarks in the fundamental representation. The inclusion of antiquarks and gluons is
straightforward. The fluid approach is based on the covariant continuity equation for the fluid four-flow
\be
\label{cont-eq}
D_\mu n^\mu  =  0 \;
 \ee
and on the equation that couples the energy-momentum tensor
$T^{\mu \nu}$ to the gauge fields 
\be \label{en-mom-eq} 
D_\mu
T^{\mu \nu} - {g \over 2}\{F_{\mu}^{\;\; \nu}, n^\mu \}= 0 \;, \ee
where the various quantities are $3 \times 3$ hermitian matrices
in color space, and we have suppressed color indices. The
covariant derivative is defined as
$$
D_{\mu} = \partial_{\mu} - ig[A_{\mu}(x),...\; ]\;,\;\;\;\;\;\;\;
$$
with $A_{\mu }=A^{\mu }_a (x) \tau^a$ for $a=1,...,8$ and $\tau^a = \lambda^a/2$, where $\lambda^a$ are the $SU(3)$
Gell-Mann matrices and therefore ${\rm Tr} (\tau^a \tau^b) = \frac 12 \delta^{ab}$.
The strength tensor appearing in Eq.(\ref{en-mom-eq}) is given by
$F_{\mu\nu}=\partial_{\mu}A_{\nu} - \partial_{\nu}A_{\mu}
-ig [A_{\mu},A_{\nu}]$.

We further assume that the four-flow and
the energy-momentum tensor have the expression valid for an ideal fluid, \ie
\be
\label{flow-id}
n^\mu(x)  = n (x) \, u^\mu(x) \;,
\ee
and
\be
\label{en-mom-id}
T^{\mu \nu}(x)  = {1 \over 2}
\big(\epsilon (x) + p (x)\big)
\big\{u^\mu (x), u^\nu (x) \big\}
- p (x) \, g^{\mu \nu}\;,
\ee
where the hydrodynamic velocity $u^\mu$,  the particle density
$n$, the energy density $\epsilon$ and the pressure $p$ are $3 \times 3$
matrices in color space.  The   quantities defined above  have in general both colorless and colored components,
as an example the particle density can be written as
\be
 n_{\alpha \beta} =  n_0 I_{\alpha \beta} + \frac 12  n_a \tau^a_{\alpha \beta}  \ ,
\ee
where $\alpha,\beta=1,2,3$ are color indices and $I$ is the identity matrix. In the following equations we will omit the color indices  not to overcharge the notation.

The color current due to the flow of the fluid  can be expressed in terms of the hydrodynamic velocity  and  the particle density as
\be
\label{hydro-current}
j^\mu(x) = -\frac{g}{2}
\Big(n u^\mu - {1 \over 3}{\rm Tr}\big[n
u^\mu  \big]\Big) \;, \ee
and it    acts  as a source term for the gauge fields in the Yang Mills equation
\be
\label{yang-mills}
D_{\mu} F^{\mu \nu}(x) = j^{\nu}(x)\; .
\ee
Thus,  we will assume that all the gauge fields that appear in the fluid equations
are  only due to the presence of a colored current in the medium.

It is worth remarking that Eqs.~(\ref{cont-eq}) and (\ref{en-mom-eq}) were derived in Ref.~\cite{Manuel:2006hg} from the collisionless transport equation obeyed by the particle distribution function. Thus, the conservation laws expressed by the Eqs.~(\ref{cont-eq}) and (\ref{en-mom-eq})  are
strictly valid on time scales shorter than the mean free path time. As an example, for quarks $n^0$ represents the quark particle density, which fulfills a conservation law that says that the change of the number of particles within a volume element is equal to the flux of particles across the surface of the volume element.  On the other hand, for times larger than the  mean free path time this conservation law   is violated by collision processes that change the particle number inside the volume element. Then, only baryon number is conserved, which 
requires the knowledge of both the quark and antiquark particle densities. 

Summarizing, for time scales shorter
than the mean free path time there are more conservation laws that at long time scales, as Eqs.~(\ref{cont-eq}) and (\ref{en-mom-eq}) indicate. Their validity for the short time phenomena that we will be studied here
is then guaranteed.

\subsection{Linearization of the chromohydrodynamic equations}

We shall consider the fluctuations of density, energy density, pressure and plasma velocity, around their stationary and colorless state described by $\bar n$, $\bar \epsilon$, $\bar p$ and
$\bar u^\mu$ respectively. In order to study such fluctuations  we linearize the chromohydrodynamic equations (\ref{cont-eq}, \ref{en-mom-eq}), assuming the fluctuations to be small. Notice that in the stationary  state the color current defined in Eq.(\ref{hydro-current})  vanishes, indeed
\be
\label{neutral-cov}
\bar j^\mu = -g \left(\bar n  \, \bar u^\mu
- {1 \over 3}{\rm Tr}\big[\bar n \, \bar u^\mu  \big]\right)
= 0 \;,
\ee
and we will assume that in the stationary state no field $F^{\mu\nu}$ is present in the system.
We define the space dependent fluctuations of the various  quantities around their  colorless values as
\be
\label{exp1}
n (x) = \bar n  + \delta n(x)
\;, \;\;\;\;\;\;\;
\epsilon (x) = \bar \epsilon + \delta \epsilon (x)
\;,
\ee
\be
\label{exp2}
p (x) = \bar p  + \delta p(x)
\;, \;\;\;\;\;\;\;
u^\mu (x) = \bar u^\mu + \delta u^\mu (x)
\;.
\ee
All the fluctuations can contain both colorless and colored components, therefore they can be decomposed as
\be
\delta n_{\alpha \beta} = \delta n_0 I_{\alpha \beta} + \frac 12 \delta n_a \tau^a_{\alpha \beta}  \ ,
\ee
where $\delta n_0 $ is the colorless fluctuation, and $\delta n_a$ the corresponding colored components.

The state described by $\bar n$, $\bar \epsilon$, $\bar p$ and
$\bar u^\mu$ is assumed to be stationary, colorless, and homogeneous on the scale of variation of
the fluctuations and therefore   we have that
\be
D^\mu \bar n  = 0 \;, \;\;\;\;\;\;\;
D^\mu \bar \epsilon  = 0 \;, \;\;\;\;\;\;\;
D^\mu \bar p  = 0 \;, \;\;\;\;\;\;\;
D^\mu \bar u^\nu  = 0 \;. \;\;\;\;\;\;\;
\ee
Moreover, since we will consider  only small deviations from the stationary state, we will assume that the following conditions are obeyed
\be
\bar n \gg \delta n \;, \;\;\;\;\;\;\;
\bar \epsilon \gg \delta \epsilon \;, \;\;\;\;\;\;\;
\bar p \gg \delta p \;, \;\;\;\;\;\;\;
\bar u^\mu \gg \delta u^\mu \;.
\ee
Actually, $\delta n$, $\delta \epsilon$, $\delta p$ and $\delta u^\mu$
should be diagonalized to be comparable to the $\bar n$, $\bar \epsilon$,
$\bar p$ and $\bar u^\mu$.

We now aim to determining the set of equations that the
fluctuations  $\delta n$, $\delta \epsilon$, $\delta p$ and $\delta u^\mu$ obey. Employing the  equations (\ref{exp1}) and (\ref{exp2}) we can derive the expression  of the  fluctuation of $n^\mu$ and of $T^{\mu \nu}$ around their colorless and homogeneous values:
\be
\label{flow-lin}
n^\mu  = \bar n \, \bar u^\mu + \bar n \, \delta u^\mu
+ \delta n \,\bar u^\mu
\;,
\ee
\be
\label{en-mom-lin}
T^{\mu \nu} = (\bar \epsilon + \bar p )
\bar u^\mu \, \bar u^\nu - \bar p  \, g^{\mu \nu}
+ (\delta \epsilon + \delta p )
\bar u^\mu \, \bar u^\nu
+ (\bar\epsilon + \bar p )
(\bar u^\mu \, \delta u^\nu + \delta u^\mu \, \bar u^\nu )
- \delta p  \, g^{\mu \nu}
\;.
\ee
Upon using these expressions in Eqs.~(\ref{cont-eq}, \ref{en-mom-eq}), we obtain that $\delta u^\mu$, $\delta p$, $\delta\epsilon$  and $\delta n$ must obey the following equations:
\ba
\label{lin-eq}
\bar n \, D_\mu \delta u^\mu
+ (D_\mu \delta n ) \, \bar u^\mu &=& 0 \;,\nonumber\\
\bar u^\mu D_\mu \delta \epsilon
+ (\bar\epsilon + \bar p ) D_\mu \delta u^\mu &=& 0 \;, \\
(\bar \epsilon + \bar p ) \bar u_\mu D^\mu  \delta u^\nu
- (D^\nu - \bar u^\nu \bar u_\mu D^\mu ) \delta p
- g \bar n  \bar u_\mu F^{\mu \nu} &=& 0\;.\nonumber
\ea

These equations do not form a closed set  and one more relation
has to be provided. In hydrodynamical treatments, one usually imposes an Equation of State (EoS) of the form
\be
\label{EoS}
  p = p(\epsilon, n) \;,
\ee
and in this way one obtains a closed set of equations. For the applications we have in mind,
we will consider that the conformal limit is reached, and further the effects of the particle
density can be ignored. Thus we will use an  Equation of State  given by 
\be
\label{EoS-RHIC}
p (x) = c_s^2 \, \epsilon(x) \,,
\ee
where  $c_s$ is the speed of sound, and in the conformal limit $c_s = 1/\sqrt{3}$. We will however leave
$c_s$ as a parameter, and  use its  value only in the  numerical analysis of the equations.

We will study separately colorless  and colored fluctuations. In order to obtain    the equation governing the colorless fluctuation we
take the trace of the Eqs.~(\ref{lin-eq}) obtaining
\ba
\label{lin-neutral}
\bar n \, \partial_\mu \delta u^\mu_0
+ (\partial_\mu \delta n_0 ) \, \bar u^\mu  =  0 \;, \nonumber
\\
\bar u^\mu \partial_\mu \delta \epsilon_0
+ (\bar\epsilon + \bar p ) \partial_\mu \delta u^\mu_0  =  0 \;,
\\
(\bar \epsilon + \bar p ) \bar u_\mu \partial^\mu  \delta u^\nu_0
- (\partial^\nu - \bar u^\nu \bar u_\mu \partial^\mu ) \delta p_0 = 0\;.\nonumber
\ea
Whereas multiplying the   Eqs.~(\ref{lin-eq})  by $\tau^a$ and taking the trace we gather the equations for colored hydrodynamical fluctuations:
\ba
\label{lin-col}
\bar n \, (D_\mu \delta u^\mu)_a
+ (D_\mu \delta n )_a \, \bar u^\mu  =  0 \;, \nonumber
\\
\bar u^\mu (D_\mu \delta \epsilon)_a
+ (\bar\epsilon + \bar p ) (D_\mu \delta u^\mu)_a  =  0 \;.
\\
(\bar \epsilon + \bar p ) \bar u_\mu (D^\mu  \delta u^\nu)_a
- ((D^\nu - \bar u^\nu \bar u_\mu D^\mu ) \delta p)_a
- g \bar n  \bar u_\mu F^{\mu \nu}_a= 0\;, \nonumber
\ea
where we have used the notation $(D_\mu X)_a \equiv \partial_\mu X^a + g f^{abc} A_\mu^b X^c$.

\section{Sound and plasma waves in the quark-gluon plasma}
\label{sound-sec}

In order to analyze the linearized chromohydrodynamical equations derived in the previous
Section we will treat separately the colorless and colored fluctuations showing that they describe the propagation of  sound and plasma waves respectively.

\subsection{Sound waves}

The colorless fluctuations of plasma velocity, energy density, pressure and density obey the Eqs.~(\ref{lin-neutral}) that in momentum space   read
\ba
\label{neutral}
\bar u^\mu k_\mu \delta n_0
+ \bar n k_\mu \delta u^\mu_0 &=& 0 \,, \nonumber\\
\bar u^\mu k_\mu \delta \epsilon_0 +
\left(\bar \epsilon + \bar p \right) k_\mu  \delta u^\mu_0
&=& 0\;,\\
i \left(\bar \epsilon + \bar p \right)
\bar u_{\mu} k^\mu  \delta u^\nu_0
+ i
(\bar u^\mu \bar u^\nu k_\mu - k^\nu )
\delta p_0 &=& 0\;, \nonumber
\ea
where  $k^\mu \equiv (\omega, {\bf k})$.
Since pressure and energy density are related by the EoS (\ref{EoS-RHIC}), their variations are not independent  and upon differentiating Eq.~(\ref{EoS-RHIC}) we obtain that
\be
\label{fluc-pressEOS}
\delta p_0 =  \left( \frac{\partial p}{ \partial \epsilon}\right)\delta \epsilon_0
\equiv   c_s^2 \,\delta \epsilon_0 \ .
\ee

Substituting this expression in Eqs.~(\ref{neutral}) we   find the following equation  for  the fluctuations of the pressure:
\be
\label{sound-wave}
\frac{1}{\bar u^\mu k_\mu} \left\{ \left(\frac{1}{c_s^2} -1 \right) (\bar u^\mu k_\mu)^2 + k^2    \right\}
\delta p_0 = 0 \ ,
\ee
while the  fluctuations of the other quantities can be expressed as
 \ba
\delta u^\nu_0 &  =  &- \frac{1}{\left(\frac {1}{c_s^2} +1 \right) \bar p} \frac{k_\mu( \bar u^\mu \bar u^\nu -
g^{\mu \nu}) }{\bar u \cdot k} \delta p_0 \ , \\
\delta n_0 &  =  &- \bar n \,\frac{k_\mu \delta u_0^\mu}{\bar u \cdot k}  \,.
\ea

In the plasma rest frame, $\bar u^\mu =(1,0,0,0)$, Eq.~(\ref{sound-wave}) simplifies to
\be
\left( \frac {1}{c_s^2} - \frac{{\bf k}^2}{\omega^2} \right) \delta p_0 = 0 \,,
\ee
which gives the  standard  expression of the sound waves in a plasma, reflecting the fact that  all colorless hydrodynamical
fluctuations propagate at the speed of sound.

\subsection{Plasma waves}

Regarding the colored components of the  fluctuations, Eqs.~(\ref{lin-col}),
we linearize in the $A$ fields as well.  The reason is that
 we will be  interested in the computation of the polarization
tensor and expanding at the first order in $A$  is sufficient. One then finds

\ba
\label{cont-eq6}
\bar u^\mu k_\mu \delta n_a
+ \bar n k_\mu \delta u^\mu_a & = & 0 \;, \\
\label{ener-den-eq6}
\bar u^\mu k_\mu \delta \epsilon_a +
\left(\bar \epsilon + \bar p \right) k_\mu  \delta u^\mu_a
& = &  0\;, \\
\label{cov-Euler6}
i \left(\bar \epsilon + \bar p \right)
\bar u_{\mu} k^\mu  \delta u^\nu_a
+ i
(\bar u^\mu \bar u^\nu k_\mu - k^\nu )
\delta p_a
- q \bar n  \bar u_{ \mu} F^{\mu \nu}_a & = & 0\;,
\ea
and  the fluctuation of the pressure and the fluctuations of the energy density can be related  using the EoS (\ref{EoS-RHIC}),  that leads to $\delta p_a = c_s^2 \delta \epsilon_a$ \cite{Manuel:2006hg},\cite{foot1}.

Equations (\ref{cont-eq6}) and (\ref{cov-Euler6}) relate  the fluctuations of the density and energy density
with the  fluctuations of the velocity
\ba
\delta n_a &  =  &- \bar n \frac{k_\mu \delta u^\mu_a}{\bar u \cdot k} \ , \\
\delta \epsilon_a &  =  &- \left(\bar \epsilon + \bar p \right) \frac{k_\mu \delta u^\mu_a}{\bar u \cdot k} \ ,
\ea

and combining these expression with Eq.~(\ref{fluc-pressEOS}) we find that   Eq.~(\ref{cov-Euler6}) can be rewritten as

\be
\label{cov-Euler7}
\Big[g^{\nu \mu} +
\frac{c^2_s}{(\bar u \cdot k)^2} \big(k^\nu
k^\mu - \bar u^\nu k^\mu (\bar u \cdot k) \big)
\Big] \delta u_{\mu, a}= + i g \frac{\bar
n}{(\bar \epsilon + \bar p) (\bar u
\cdot k)} \bar u_{ \mu} F^{\mu \nu}_a \;.
 \ee
The inverse of the operator on the left hand side of this equation can be determined observing that
\be
\Big[g^{\nu \mu} + \frac{c^2_s}{(\bar u \cdot k)^2}
\big(k^\nu k^\mu - \bar u^\nu k^\mu (\bar u \cdot k)
\big) \Big] \; \Big[g_{\sigma \nu} - {\cal T}(k) \big(k_\sigma
k_\nu - \bar u_{ \sigma} k_\nu (\bar u \cdot k) \big)
\Big] = g_\sigma^{\;\;\mu} \;,
\ee
where
 \be {\cal T}(k) =
 - \frac{1}{ k^2 + (\frac{1}{c^2_s}
-1) (\bar u \cdot k)^2} \,.
\ee

Then one can solve Eq.~(\ref{cov-Euler7})  and the fluctuation of the hydrodynamic velocity takes the form
\be
\delta u_{ \sigma, a} =  i g
\frac{\bar n}{(\bar \epsilon + \bar
p) (\bar u \cdot k)}
\Big[g_{\sigma \nu} +  {\cal T}(k)
\big(k_\sigma k_\nu - \bar u_{\sigma} k_\nu
(\bar u \cdot k) \big) \Big]
\bar u_{\mu} F^{\mu \nu}_a \,.
\ee

The   fluctuation of the current induced by the fluctuation of the density and of the hydrodynamic velocity can be derived from
Eq.(\ref{hydro-current}) and is given by
\be
\label{curr-lin}
\delta j^\mu_a = -\frac{g}{2} \left(\bar
n \,\delta u^\mu_a + \delta n_a \, \bar
u^\mu  \right)\;.
\ee
Such fluctuations are related in linear response theory to fluctuations of the gauge fields via
\be
\delta j^\mu_a(k) = -
\Pi^{\mu \nu}_{ab} (k) A_{\nu,b}(k) \;, \ee
where $\Pi^{\mu \nu}_{ab}$ is the polarization tensor.
Considering that the linearized strength tensor equals $F^{\mu \nu}_a(k) = - ik^\mu A^\nu_a(k) +
ik^\nu A^\mu_a(k)$,  and upon substituting the values of the fluctuations of density and energy density in Eq.(\ref{curr-lin})
we obtain that
\ba
\label{Pi2}
\nonumber
\Pi^{\mu \nu}_{ab}(k) = - \delta_{ab} \frac{g^2}{2}
\frac{\bar n^2}{(\bar \epsilon + \bar p)}
\frac{1}{(\bar u \cdot k)^2}
&\Big[&
(\bar u \cdot k)
(k^\mu \bar u^\nu + k^\nu \bar u^\mu )
- (\bar u \cdot k)^2 g^{\mu \nu}
- k^2  \bar u^\mu \bar u^\nu \\[2mm]
&-& \frac{1}{ k^2 + (\frac{1}{c^2_s} -1) (\bar u \cdot
k)^2} \Big((\bar u \cdot k)k^2 (k^\mu \bar u^\nu +
k^\nu \bar u^\mu ) - (\bar u \cdot k)^2 k^\mu k^\nu
- k^4  \bar u^\mu \bar u^\nu \Big)
 \: \Big]\;.
\ea
One can easily check that the polarization tensor (\ref{Pi2}) is symmetric
and transverse
($k_\mu \Pi^{\mu \nu}(k) = 0$).

In the plasma rest frame, $\bar u^\mu =(1,0,0,0)$, the polarization tensor simplifies
considerably and its components are given by
\ba
\Pi^{00}_{ab} (\omega, {\bf k}) & = &- \delta_{ab}\, \omega^2_p \frac{k^2}{\omega^2 - c_s^2 {\bf k}^2}  \ ,\nonumber\\
\Pi^{0i}_{ab} (\omega, {\bf k}) & = & \Pi^{i0}_{ab} (\omega, {\bf k}) = - \delta_{ab}\, \omega^2_p \frac{\omega k^i}{\omega^2 - c_s^2 {\bf k}^2} \ ,\label{pimunu-rest}
 \\
\Pi^{ij}_{ab} (\omega, {\bf k}) & = &- \delta_{ab}\, \omega^2_p \left(\delta^{ij} + \frac{c_s^2 k^i k^j}{\omega^2 - c_s^2 {\bf k}^2}\right) \nonumber \ ,
\ea
where
\be
\omega^2_p = \frac{g^2}{2}
\frac{\bar n^2}{(\bar \epsilon + \bar p)}
\ee
is the plasma frequency.

At the linear order, the Fourier transformed chromodynamic field $A^{\mu}(k)$ satisfies the equation of motion
\be
\label{eq-A}
\Big[ k^2 g^{\mu \nu} -k^{\mu} k^{\nu} - \Pi^{\mu \nu}(k) \Big]
A_{\nu}(k) = 0 \;,
\ee
where we have  dropped the color indices $a,b$.

In order to determine the dispersion law for the gauge fields we define the dielectric tensor $\varepsilon^{ij}(k)$
according to
\be
\varepsilon^{ij}(k) = \delta^{ij} + {1 \over \omega^2} \Pi^{ij}(k) \;,
\label{eps-pi}
\ee
that in the plasma rest frame  is given by
 \be
\label{epsilon-p}
\varepsilon^{ij}_{\rm p}(\omega,{\bf k}) =
\Big(1 - \frac{\omega_{\rm p}^2}{\omega^2} \Big) \delta^{ij}
- \frac{\omega_{\rm p}^2}{\omega^2} \frac{c_s^2 k^i k^j}{\omega^2 - c_s^2 {\bf k}^2}
\,.
\ee

The dispersion equation can now
be determined by computing the determinant of the matrix
\be
\label{M-matrix}
M^{ij}={\bf k}^2 \delta^{ij} -k^i  k^j
- \omega^2 \varepsilon^{ij}(k)\, ,
\ee
and solving the equation
\be
\label{dispersion-g}
{\rm det}\Big[M^{ij}\Big]  = 0 \,,
\ee
where we assumed the Coulomb
gauge, ${\bf k} \cdot {\bf A}(k)=0$, and $A^0 = 0$, meaning  that
${\bf E} = i\omega {\bf A}$.

Taking the wave vector in a given direction, say ${\bf k} = (k,0,0)$, one finds the
following solutions to the dispersion relation
\ba
\label{longitudinal-plasma}
\omega^2 & = & \omega_p^2 + c_s^2 k^2 \ , \\
\label{transverse-plasma}
\omega^2 & = &\omega_p^2 + k^2  \ ,
\ea
which correspond to longitudinal and transverse modes, respectively. 
Notice also that both modes are not damped.
The reason is that we have neglected the dissipative terms in the hydrodynamical equations, assuming that the colored plasma on the time scale of interest to the present analysis can be approximated as an ideal fluid.
In order to consider the damping of color fluctuations in the hydrodynamical limit one should include
transport coefficients, such as color conductivity or color diffusion.

Let us compare the dispersion laws of the longitudinal and transverse
modes obtained above with the corresponding results in  kinetic theory. Using the hard thermal loop (HTL) effective theory one can determine the  polarization tensor for the gauge fields. The  dispersion laws in the limit $k \ll \omega$ for the longitudinal and transverse modes are given by \cite{Lebellac}
\ba
\label{longitudinal-plasma-HTL}
\omega^2 & = & \omega_p^2 + \frac 35 k^2 \ , \\
\label{transverse-plasma-HTL}
\omega^2 & = &\omega_p^2 + \frac 65 k^2  \ ,
\ea
respectively. The chromohydrodynamical approach reproduces the short time physics of kinetic theory in the domain
$k \ll \omega$. There is a numerical discrepancy in the $k^2$ terms when comparing the two set
of dispersion laws. The origin of such a discrepancy is that in order to go from transport theory
to a fluid approach, one has to impose a relation between the energy and the pressure of the system that
allows to discard all the higher momenta moments that are contained in the particle distribution
function. Such an approximation, typical of an hydrodynamical approach, does not affect in any
profound manner the physics described in the domain of applicability of the formalism.

\section{A relativistic jet traversing the plasma}
\label{plasma+jet-sec}

We  shall now consider a jet of particles propagating across the plasma at a relativistic velocity $v=|\bf v|$. We  assume that both the jet and the plasma  are initially in equilibrium  but  jet  and plasma are not in equilibrium with each other. The system formed by the plasma and the jet is one specific example of the so-called two stream systems, which have been widely studied in magnetohydrodynamics, and extended to the chromohydrodynamical case in  Ref.~\cite{Manuel:2006hg}.

We are not  concerned with the colorless hydrodynamical
fluctuations, which have been studied elsewhere \cite{Stoecker:2004qu,Casalderrey-Solana:2004qm}.
We will instead examine the evolution of the colored fluctuations at short time scales, where the equations describing such fluctuations can be linearized.

We consider the frame where the plasma is at rest. The velocity, as well as the pressure, energy density and particle density of the  jet and of the plasma  will be different and will be labeled in a different way.
Since  the plasma is  at rest,  $\bar u^{\mu}_{\rm p} = (1,0,0,0)$, whereas the velocity of the jet  will  have the general expression
$\bar u^{\mu}_{\rm jet} = \gamma (1,{\bf v})$, with $\gamma=1/\sqrt{1-v^2}$.
The polarization tensor associated to the jet can be computed in an analogous way to that of the plasma and will have  the expression reported in Eq.~(\ref{Pi2}) with $\bar u^{\mu} \to \bar u^{\mu}_{\rm jet}$. In the limit where $\gamma \gg 1$, it is possible to neglect the second term in the square bracket on the right hand side of Eq.~(\ref{Pi2}). Such an approximation  corresponds to solving the fluid equations neglecting the effect of the pressure gradients. One can also see that this approximation corresponds to considering that the distribution function of the constituents of the jet is of
the form (see \cite{Manuel:2006hg} for a discussion of this approximation)
\be
\label{tsunami}
f(p) = \bar n \, \bar u^0 \;
\delta^{(3)}\Big({\bf p}
- \frac{\bar \epsilon + \bar p}
{\bar n} \, {\bf u} \Big) \; ,
\ee
which describes fast moving particles, with a non-thermal distribution. 
 
Employing Eq.~(\ref{eps-pi}) we obtain  the following expression of the dielectric tensor for the jet:
\be
\label{epsilon-jet}
\varepsilon^{ij}_{\rm jet}(\omega,{\bf k}) =
\Big(1 - \frac{\omega_{\rm jet}^2}{\omega^2} \Big) \delta^{ij}
- \frac{\omega_{\rm jet}^2}{\omega^2}
\bigg( \frac{ v^i k^j + v^j k^i}
{\omega - {\bf k} \cdot  {\bf v}}
- \frac{(\omega^2 - {\bf k}^2) v^i  v^j}
{(\omega - {\bf k} \cdot {\bf v} )^2} \bigg) \;,
\ee
where
\be
\omega^2_{\rm jet} = \frac{g^2}{2}
\frac{\bar n^2_{\rm jet}}{(\bar \epsilon_{\rm jet} + \bar p_{\rm jet})}
\ee
is the plasma frequency squared of the jet.

In order to obtain the dielectric tensor of the system composed by the plasma and the jet  we note that the total polarization tensor of the system is given  by the sum of the two polarization tensors in linear response theory
\be
\Pi_{\rm t}^{\mu\nu} = \Pi_{\rm p}^{\mu\nu} + \Pi_{\rm jet}^{\mu\nu}\,,
\ee
where the expression of the components of $\Pi_{\rm p}^{\mu\nu}$  is given in Eq.~(\ref{pimunu-rest}).
The  dielectric tensor of the total system turns out to be
\be
\label{total-dielectric}
\varepsilon^{ij}_{\rm t}(\omega,{\bf k}) =
\Big(1 - \frac{\omega_{\rm t}^2}{\omega^2} \Big) \delta^{ij}
- \frac{\omega_{\rm p}^2}{\omega^2} \frac{c_s^2 k^i k^j}{\omega^2 - c_s^2 {\bf k}^2} -\frac{\omega_{\rm jet}^2}{\omega^2}
\bigg( \frac{ v^i k^j + v^j k^i}
{\omega - {\bf k} \cdot  {\bf v}}
- \frac{(\omega^2 - {\bf k}^2) v^i  v^j}
{(\omega - {\bf k} \cdot {\bf v} )^2} \bigg)\,,
\ee
where
\be
\omega_{\rm t}^2 = \omega_{\rm p}^2+\omega_{\rm jet}^2  \, .
\ee
The dispersion laws of the collective modes of the system composed by the plasma and the jet can now be determined solving  the equation
\be
\label{dispersion-T}
{\rm det}\Big[M^{ij}_t\Big]  \equiv {\rm det}\Big[ {\bf k}^2 \delta^{ij} -k^i  k^j
- \omega^2 \varepsilon^{ij}_{\rm t}(k)  \Big]  = 0 \,.
\ee
The solutions of this equation depend on ${|\bf k|}$, ${|\bf v|}$,
$\cos\theta={\bf \hat k \cdot \hat v}$, $\omega_{\rm p}$, $\omega_{\rm jet}$ and $c_s$.
We find that Eq.~(\ref{dispersion-T}) admits eight solutions, however two of them are trivially given by $\omega =0$ and
six  of these solutions are given by the roots of a polynomial of the sixth order.
In order to  simplify the expression of the dispersion law and to study the dependence of the unstable collective modes on the   values of the various parameters,  we define the  dimensionless quantities
\be
\label{variables}
x = \frac{\omega}{\omega_{\rm t}} \ , \qquad y = \frac{k}{\omega_{\rm t}} \ ,
\qquad b=\frac{\omega_{\rm jet}^2}{\omega_{\rm t}^2} \ ,
\ee
meaning that we will  measure momenta and energies in units of  $\omega_{\rm t}$.
Since the plasma frequency of the jet is unknown we will treat $b$ as a parameter and we will analyze values of $b \ll 1$ corresponding to $\omega_{\rm jet} \ll  \omega_{\rm p}$.

We anticipate that we find that five of the dispersion laws are always stable, whereas one is  unstable in a certain
range of values of the parameters. In particular we find that for values
of  $v$ smaller than the speed of sound of the plasma, $c_s$, all the modes are stable independent of
the values of the remaining parameters.  Let us also note that when ${\bf k}$ and ${\bf v}$ are not
parallel, it is not possible to decompose the dielectric function (\ref{total-dielectric}) with
only transverse and longitudinal projectors of ${\bf k}$. We will then not refer to transverse or
longitudinal modes in that case.

In the following Subsections, we will analyze the dispersion relations of the collective modes for different
 orientations of ${\bf k}$ and  ${\bf v}$.

\subsection{{\bf k} parallel to {\bf v}  }\label{longsec}

Here we consider the case corresponding to  ${\bf k} \parallel {\bf v}$, and choose
 ${\bf k} = (0,0,k)$ and ${\bf v} =( 0,0,v)$.
This case can be easily analyzed because for this orientation of $\bf k$ and $\bf v$  the dielectric tensor (\ref{total-dielectric}) and the matrix $M^{ij}_t$ defined in Eq.(\ref{dispersion-T}) are  diagonal and the corresponding
dispersion equation factorizes. Moreover, since  $\bf k$ is parallel to $\bf v$ it is possible to define transverse and longitudinal modes with respect to the orientation of these vectors. We find  two stable transverse modes with dispersion law
\be
\label{transverse1}
\omega^2  = \omega_t^2 + k^2 \, .
\ee
Notice that this dispersion law is analogous to the dispersion law that we obtained in Eq.(\ref{transverse-plasma}) that describes the propagation of  the transverse mode  in a a plasma without a jet. The only difference is that  the plasma frequency $\omega_{\rm p}$ has been replaced  with  $\omega_{\rm t}$. Therefore the only effect on the transverse modes due to the   presence of the jet is to change the plasma frequency of the mode.

Regarding the longitudinal modes, they  are given by the solution
of the following equation:
 \be \label{long-unstable} \omega^2 -
\omega_{\rm t}^2 - \omega_{\rm p}^2 \frac{c_s^2 k^2}{\omega^2 -
c_s^2  k^2} -\omega_{\rm jet}^2
 \frac{ 2 v k \omega - v^2\omega^2 - v^2 k^2}
{(\omega -  k v )^2} = 0
\,.\ee
In terms of the dimensionless variables defined in Eq.~(\ref{variables}),  the equation for the longitudinal modes (\ref{long-unstable})  can be   written as
\be
\label{long-general}
x^2 \Big[ x^4 - 2 v y x^3 + x^2 \left( -1 + b v^2 + y^2 (v^2 - c_s^2) \right) + 2 v y x (1 -b + c_s^2 y^2)
- y^2 \left(b v^2 (c_s^2 - 1 ) - c_s^2 b +    v^2(c_s^2 y^2 + 1 ) \right) \Big] = 0 .
\ee
This equation has two trivial solutions $x=0$,   and four solutions corresponding to the roots of a quartic equation.  As already mentioned we find that at most
one mode is unstable. In the present case one of the longitudinal modes is unstable for $c_s< v < 1$.  Since the cases where $v=c_s$ and  $v \to 1$ will play a special role for these modes, let us first consider the solutions of Eq.~(\ref{long-unstable}) for these two values of the velocity.

In the limit   $v \to 1$, Eq.~(\ref{long-unstable}) simplifies to
\be
\label{long-unstablev=1}
\omega^2 - \omega_{\rm t}^2
- \omega_{\rm p}^2 \frac{c_s^2 k^2}{\omega^2 - c_s^2  k^2} +\omega_{\rm jet}^2
 = \omega^2 - \omega_{\rm p}^2
- \omega_{\rm p}^2 \frac{c_s^2 k^2}{\omega^2 - c_s^2 k^2} = 0
\,,\ee
with solution, $\omega^2 =0$ and
\be
\omega^2  = \omega_p^2 + c_s^2 k^2 \ .
\ee

This solution is the same that we have obtained for a plasma
without a jet in Eq.(\ref{longitudinal-plasma}). Therefore in the
limit $v \to 1$, the longitudinal mode is stable and we see that a
jet of particles moving at the speed of light can only affect the
transverse modes , but they do not affect at
all the longitudinal modes. This is related to the dimensional
contraction which occurs in the eikonal limit
\cite{Jackiw:1991ck}, when it can be shown that the strength field
tensor caused by a massless particle that moves at the speed of
light is on the plane transverse to its motion.

In the case $v =c_s$ it is possible to find one simple analytical  solution  (apart from the trivial
solution $\omega^2 = 0$ and Eq.~(\ref{transverse1})) which is given by
\be
 \omega = c_s k \ ,
\ee
whereas  three more solutions  correspond  to the roots of
\be
x^3 - c_s x^2 y + c_s y (1 + b(-2 + c_s^2)+ x(-1+ c_s^2 (b - y^2)) +
          c_s^2 y^2)  = 0 \,.
\ee
The discriminant of this equation is negative for $b < 1$ meaning that  the three solutions of this equation are always real
and the corresponding modes stable.

\begin{figure}[!th]
\includegraphics[width=3.5in,angle=-0]{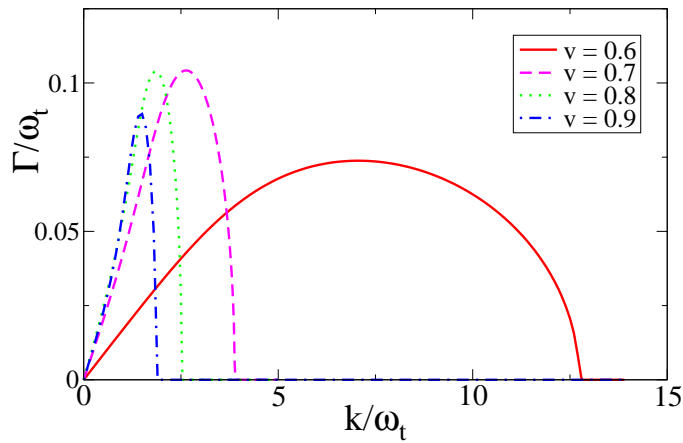}
\caption{(Color online) Imaginary part of the dispersion law of  the unstable longitudinal mode for the system composed by a plasma and a jet
as a function of  $k/\omega_{\rm t}$. Here the momentum of the unstable mode is parallel to the velocity of the jet $\bf v$, $b=\omega_{\rm jet}^2/\omega_{\rm t}^2=0.02$  and the four different lines correspond to   different values of
 the velocity of the jet $|{\bf v}|$. \vspace{1cm}} \label{Parafig1}
\end{figure}

We now turn to a  numerical study of the non-trivial  solutions of Eq.~(\ref{long-general})
as a function of the parameters $b$ and $v$. We find a  solution with a positive imaginary part for any non-vanishing value   of $b$  and for $v$ larger than $c_s$. Such a solution corresponds to the unstable longitudinal collective mode of the system.
In Fig.~\ref{Parafig1} we have reported the
plot of the imaginary part of the frequency, $\Gamma$, of the  this longitudinal mode as  a function of $k/\omega_t$,  for $b=0.02$ and for four different values of the velocity of the jet: $v=0.6$ full (red online) line; $v=0.7$, dashed (magenta online) line; $v=0.8$, full (green online) line; $v=0.9$, dot-dashed (blue online) line. For
any value of the velocity smaller or equal the speed of sound we numerically find $\Gamma=0$. For  values of $v$ larger than the speed of sound and less than $1$ this collective mode becomes unstable  in a  range of values of the momentum $k<k_{\rm max}$.
For small values of $k$, $\Gamma$ increases with increasing momentum and reaches  a peak. Then it decreases and at $k_{\rm max}$ becomes zero.  As can be seen in Fig.~\ref{Parafig1} for $v > c_s$, the  value of $k_{\rm max}$ decreases with increasing  velocity.

\begin{figure}[!th]
\includegraphics[width=3.5in,angle=-0]{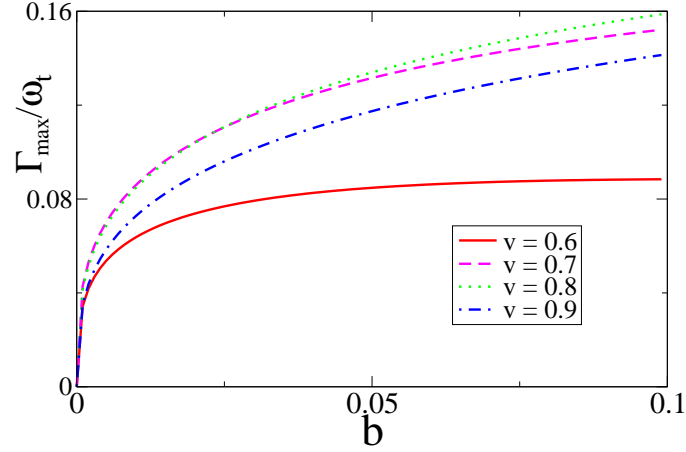}
\caption{(Color online) Variations of the largest value of  the imaginary part of the dispersion law of the unstable longitudinal mode for the system composed by a plasma and a jet in the case $\bf k \parallel v$ as a function of $b$  for four different values of the velocity of the jet,  $|{\bf v}|$. \vspace{1cm}} \label{Parafig2}
\end{figure}

The  value of  $\Gamma$ at the peak,  that we will indicate  with $\Gamma_{\rm max}$, depends on $v$ and $b$.
 In Fig.~\ref{Parafig2} we present the plot
of $\Gamma_{\rm max}$ as a function of $b$ for four values of the jet velocity: $v=0.6$ full (red online) line; $v=0.7$, dashed (magenta online) line; $v=0.8$, full (green online) line; $v=0.9$, dot-dashed (blue online) line. Note that independent of the value of $v$, for $b=0$, we obtain that $\Gamma=0$. In this case the plasma frequency of the jet is zero and corresponds to the case where the system consists of the plasma only. Indeed, for $b=0$, the dielectric tensor (\ref{total-dielectric}) does not have any contribution from the jet and therefore all the collective modes are stable, as shown in the previous Section.
 With increasing values  of $b$, \ie of the plasma frequency of the jet, $\Gamma_{\rm max}$ increases, meaning that to larger values of  the density of the jet (for fixed values of $g$ and $\bar \epsilon_{\rm jet} + \bar p_{\rm jet}$) correspond larger instabilities.
Moreover, for values of $b \lesssim 0.01$ the value of $\Gamma_{\rm max}$ increases quickly with increasing $b$ whereas for larger values of $b$, $\Gamma_{\rm max}$ becomes less sensitive to the actual value of $b$. In particular for  $v=0.6$, $\Gamma_{\rm max}$ saturates at $\sim 0.09 \,\omega_{\rm t}$.
As a function of the velocity we find that for a given value of $b$  the value of $\Gamma_{\rm max}$ increases with increasing velocity for $c_s<v \lesssim 0.7$.  For $v \sim 0.7-0.8$ the value of $\Gamma_{\rm max}$ reaches its maximum value and then for larger values of the velocity  decreases  and eventually in the ultrarelativistic case
$v \to 1$ we find that  the frequency  of the longitudinal mode becomes real for all the values of $k$.

\subsection{ {\bf k} orthogonal to {\bf v}}\label{orthosec}

We now consider the case corresponding to  ${\bf k}$ orthogonal to ${\bf v}$,
choosing ${\bf k} = (k,0,0)$ and ${\bf v} =( 0,0,v)$. In this case the matrix  $M^{ij}_t$ defined in Eq.(\ref{dispersion-T}) is block  diagonal. One block is one-dimensional and corresponds to the component  $M^{yy}_t$. The other block is two-dimensional and corresponds to the components of $M^{ij}_t$ with indices $i,j=x,z$. Therefore the
dispersion equation (\ref{dispersion-T}) factorizes into two equations.
Regarding the one-dimensional block it describes a mode orthogonal to both $\bf k$ and $\bf v$. This mode is stable and has dispersion law
\be
\omega^2  = \omega_t^2 + k^2  \ .
\ee
This dispersion law is analogous to the dispersion law for the transverse mode for
a plasma without a jet, given in Eq.(\ref{transverse-plasma}), with the plasma frequency $\omega_{\rm p}$ replaced  with  $\omega_{\rm t}$.
The remaining collective modes are solutions of the equation
\be
\label{trans-unstable}
\left(\omega^2 - \omega_{\rm t}^2
- \omega_{\rm p}^2 \frac{c_s^2 k^2}{\omega^2 - c_s^2  k^2}\right) \left(\omega^2 - k^2 - \omega^2_t
-\omega_{\rm jet}^2 \frac{v^2 (k^2 - \omega^2)}{\omega^2} \right) -
\omega_{\rm jet}^4 \frac{v^2 k^2}{\omega^2}
= 0
\,.
\ee
Using the variables defined in Eq.~(\ref{variables}), one can rewrite this equation as
\be
\label{x-eq}
x^2 \left(x^2 - 1 - c^2_s y^2 \right) \left(x^2 -1- y^2 + v^2 b \right) + x^2 y^2 b (c_s^2 - v^2)
+ b y^2 \left( v^2 -c_s^2 + c_s^2 y^2 (v^2-1) +c_s^2 v^2 b\right) - b^2 v^2 y^2= 0 \ ,
\ee
where we have already factored out the trivial solution $x^2 = 0$.

\begin{figure}[!th]
\includegraphics[width=3.5in,angle=-0]{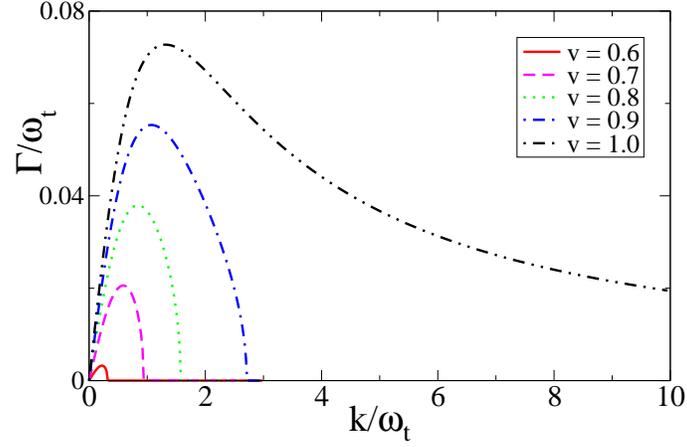}
\caption{(Color online) Imaginary part of the dispersion law of the unstable  mode for the system composed by a plasma and a jet in the case where the momentum of the collective mode, $\bf k$, is orthogonal to the velocity of the jet, $\bf v$, as a function of  the momentum $k / \omega_{\rm t}$ and for  $b=\omega_{\rm jet}^2/\omega_{\rm t}^2=0.02$.  The  five different lines correspond to   different values of the velocity $|{\bf v}|$. \vspace{1cm}.} \label{Orthofig1}
\end{figure}

\begin{figure}[!th]
\includegraphics[width=3.5in,angle=-0]{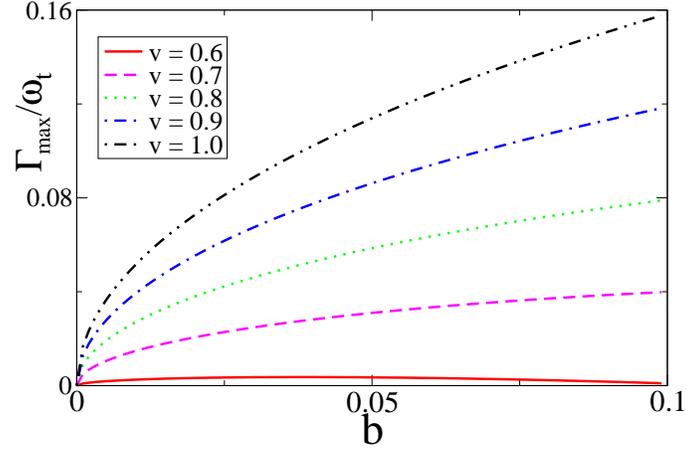}
\caption{(Color online)   Largest value of the  imaginary part of the dispersion law of  the unstable mode  for the system composed by a plasma and a jet in the case where the momentum of the collective mode, $\bf k$, is orthogonal to the velocity of the jet, $\bf v$, as a function of $b=\omega_{\rm jet}^2/\omega_{\rm t}^2$.
The  five different lines correspond to   different values of the velocity $|{\bf v}|$.  \vspace{1cm}} \label{Orthofig2}
\end{figure}

Also in this case we find that for $v \lesssim c_s$ no mode is unstable and for  $v > c_s$ only one mode is unstable in a certain range of values of the momentum. We have numerically studied the solutions of Eq~(\ref{x-eq}) and we have reported the results in
Fig.~\ref{Orthofig1} and Fig.~\ref{Orthofig2}.  In Fig.~\ref{Orthofig1}  it is shown the plot of the imaginary part of  $\omega$ for $b=0.02$,
as a function of $k$ for five different values of the jet velocity: $v=0.6$, full (red online) line; $v=0.7$, dashed (magenta online) line; $v=0.8$, full (green online) line; $v=0.9$, dot-dashed (blue online) line; $v=1$ dot-dot-dashed (black) line.

The largest value of $k$ where the orthogonal mode is unstable is  a monotonic increasing  function of $v$ and it diverges for $v\to 1$. Also  the
value of the imaginary part of $\Gamma$ at the peak, $\Gamma_{\rm max}$,  increase with increasing  $v$, but for $v \to 1$ reaches a maximum  finite value.
The behavior of the  largest value of the imaginary part of the frequency, $\Gamma_{\rm max}$, as a function of $b$, for various values
of the velocity, is reported in Fig.~\ref{Orthofig2}. With increasing values of
$b$, \ie with increasing value of the plasma frequency of the jet, the value of
$\Gamma_{\rm max}$  increases. Moreover, for a given value of $b$, the larger the velocity of the jet, the  larger is the  value  of  $\Gamma_{\rm max}$.

\subsection{Arbitrary angles}\label{obliqsec}
In the previous two Subsections we have analyzed the behavior
of the unstable mode of the plasma for  two values of the angle $\theta$ between $\bf k$ and  $\bf v$. In  particular in Section \ref{longsec} we have considered $\theta=0$, corresponding to modes with $\bf k \parallel v $ and in Section \ref{orthosec}  we have taken $\theta=\pi/2$, corresponding to  modes with $\bf k \perp v $.   A common feature
of the two cases is that the value of $\Gamma_{\rm max}$ (corresponding
to the largest value of the imaginary part of the frequency as a function of $|\bf k|$), increases with increasing values of the plasma frequency of the jet.
Regarding the dependence of $\Gamma_{\rm max}$ on the value of the velocity of the jet
$v$ one can see, comparing  the plots in Fig.~ \ref{Orthofig2} with those in Fig.~ \ref{Parafig2}, that for ultrarelativistic jet the modes with $\bf k \perp v$ dominate the
modes with $\bf k \parallel v$.    On the other hand, for velocity larger than the speed of sound,
but $v \lesssim 0.9$  the longitudinal modes are dominant.

In this Subsection we analyze the general case, of an arbitrary angle $\theta$
and we will determine at which angle corresponds the most unstable mode as a
function of $v$ and $b$. For definitiveness we choose ${\bf v} = (0,0,v) $  and ${\bf k} = (0,k \sin\theta,k \cos\theta)$.

The analysis of the dispersion laws for the modes with $\bf k \parallel v $  and  $\bf k \perp v $
 was simplified by the fact that in both cases the matrix $M^{ij}_t$ defined in Eq.~(\ref{dispersion-T}) becomes block-diagonal and the corresponding  equation for the  dispersion laws factorizes. However, for arbitrary angle $\theta$  the matrix $M^{ij}_t$  is not block-diagonal. Then we will rely on a numerical solution of Eq.~(\ref{dispersion-T}).
Employing the definitions in Eq.(\ref{variables}) we obtain (apart from the two trivial solutions corresponding to  $\omega=0$), that the  six dispersion laws can be obtained from the roots of the equation
\ba
0 & = &2x^6 + x^4(-4 + 2bv^2 + (-2 - 2c_s^2 + v^2)y^2) + x^2(2 - 2bv^2 - (-2 - 2(1 + b)c_s^2 + (2 + b + 2bc_s^2)v^2)y^2  \nonumber \\  && - (v^2 +c_s^2(-2 + v^2))y^4) + by^2(-(v^2(b + y^2)) + c_s^2(-2(1 + y^2) + v^2(1 + b + 2y^2)))   \nonumber  \\ & & + vy(v(1 + c_s^2y^2)(y + y^3) - 4x(-1 + x^2 - y^2)(-1 + b + x^2 - c_s^2y^2)\cos(\theta)   \nonumber  \\  & & + vy(-(b^2(-1 + c_s^2)) + b(-2 + c_s^2 + x^2 - y^2) + (-1 + x^2 - y^2)(-1 + x^2 -c_s^2y^2))\cos(2\theta)) \,.
\ea

\begin{figure}[!th]
\includegraphics[width=2.3in,angle=-0]{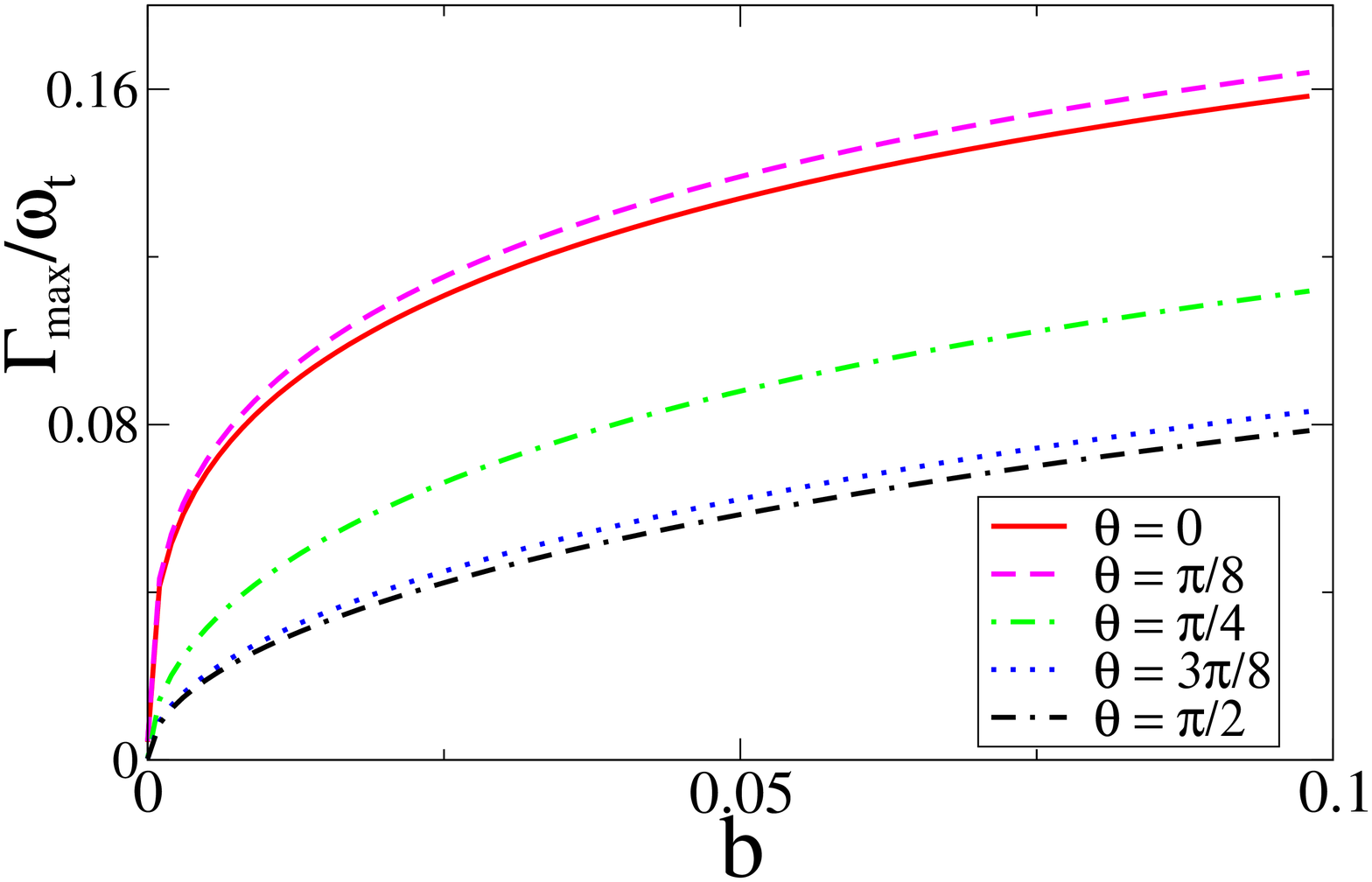}
\includegraphics[width=2.3in,angle=-0]{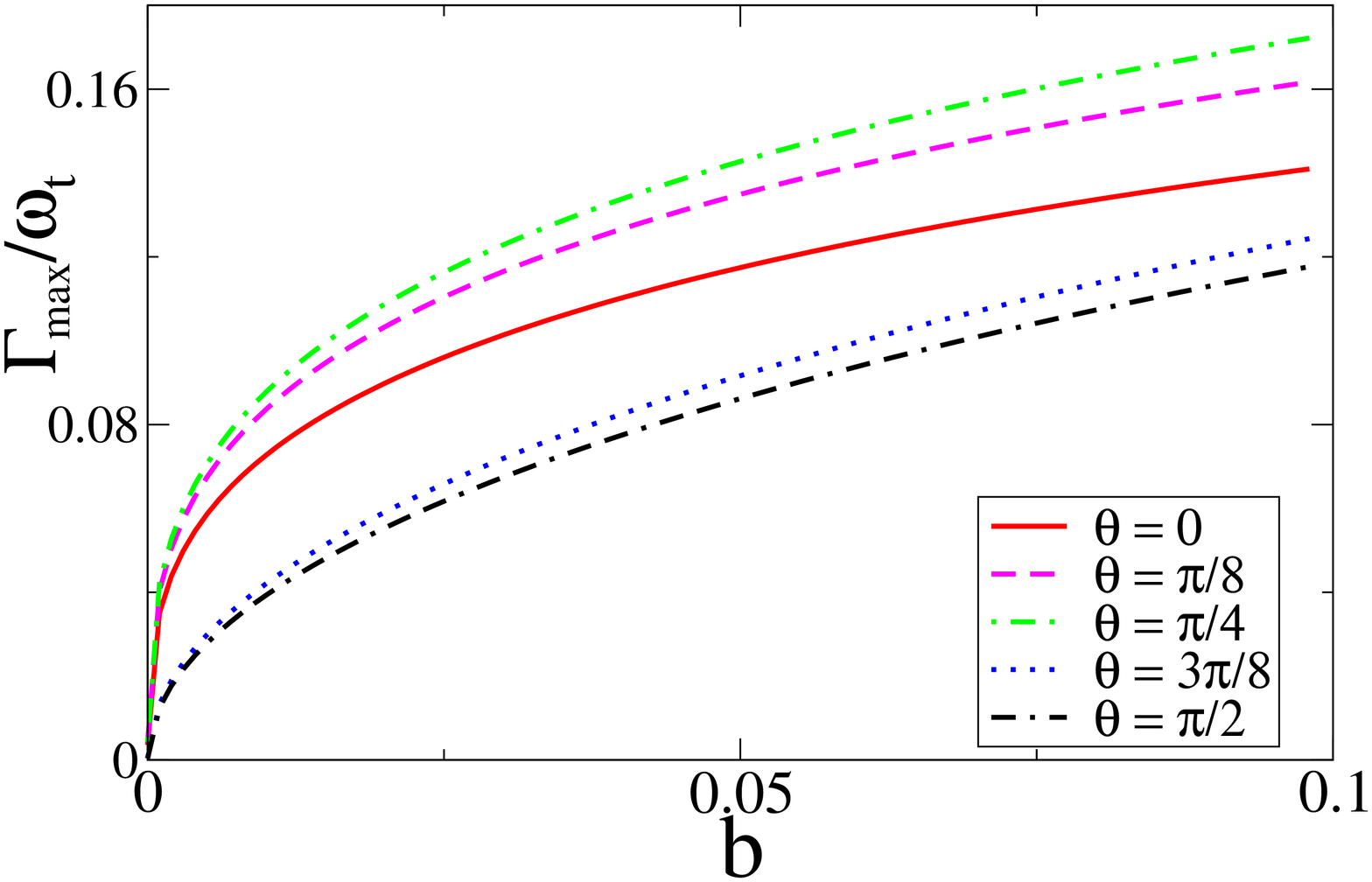}
\includegraphics[width=2.3in,angle=-0]{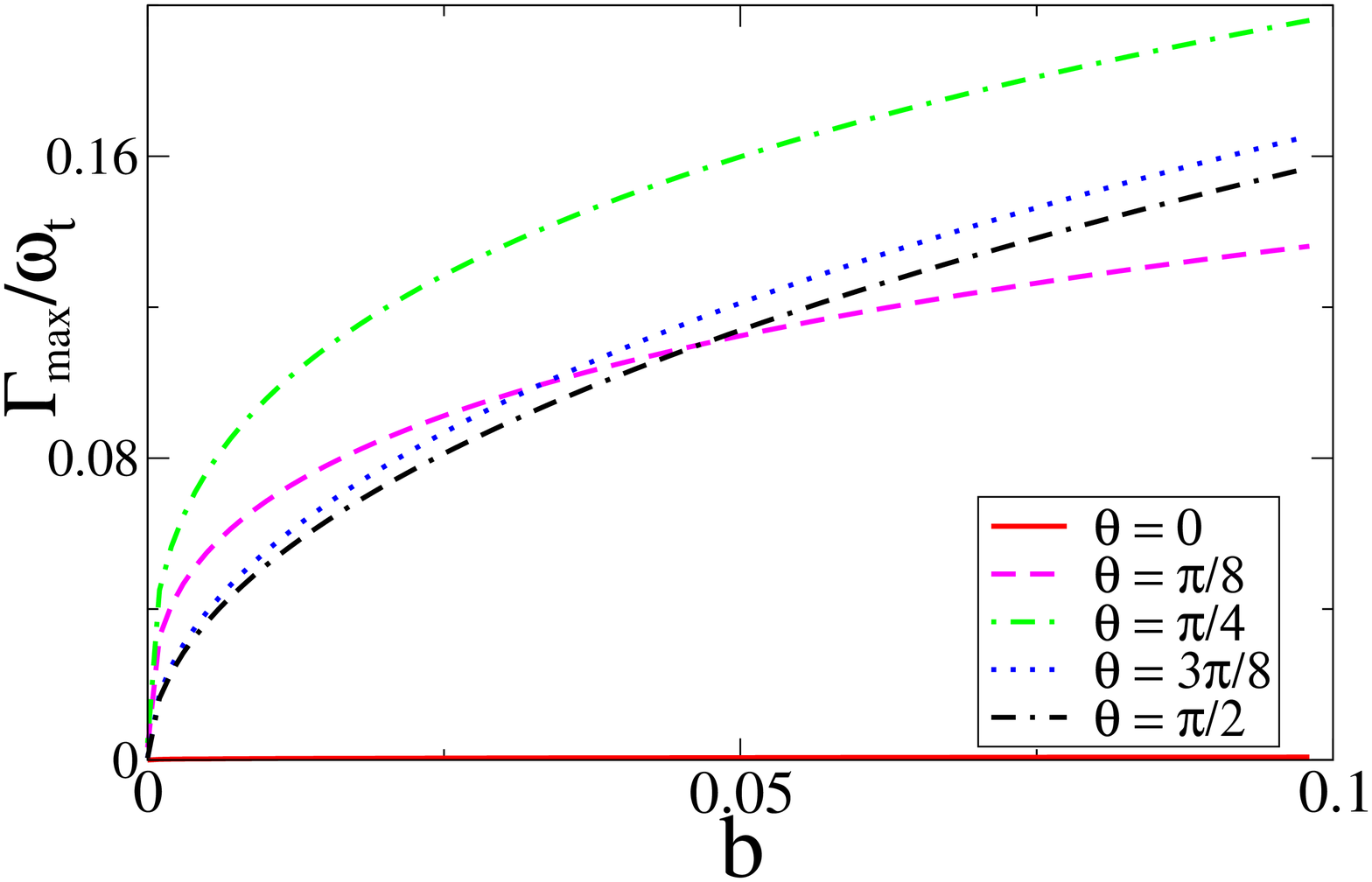}
\caption{(Color online) Largest value of the imaginary part of the dispersion law for the unstable mode as a function of $b$ for three different values of  the velocity of the jet ${|\bf v|}$ and five different angles between $\bf k$ and $\bf v$. The left panel corresponds to $v=0.8$ and the dominant unstable modes correspond to small angles $\theta$, \ie to modes that are almost collinear with the jet. The central panel corresponds to $v=0.9$ and the right panel to $v \to 1$.  In both cases the modes corresponding to $\theta \sim \pi/4 $ are the dominant one. In the limit $v \to 1$ modes with momentum collinear with the velocity of the jet  are suppressed and the mode with $\theta=0$ vanishes.  \vspace{1cm}} \label{obliquefig1}
\end{figure}

Of the six solutions of this equation only one corresponds to an unstable mode in a certain range of parameters.
In particular, for any value of $\theta$ we find an unstable mode only if $v>c_s$ and if the momentum is smaller than a certain value $k_{\rm max}$, where $k_{\rm max}$ depends on $\theta$, $v$ and $b$.
The largest value of the imaginary part $\Gamma_{\rm max}$ depends on $\theta$, $v$ and $b$ as well and
in Fig.~\ref{obliquefig1} we show three plots of $\Gamma_{\rm max}$  as a function of $b$ for various values of $\theta$ and $v$.
In each plot the  full (red online) line corresponds to $\theta =0$, the dashed (magenta online) line corresponds to $\theta =\pi/8$, the dot-dashed (green online) line corresponds to $\theta=\pi/4$, the dotted (blue online) line to $\theta = 3\pi/8$ and the dot-dashed-dashed (black) line to  $\theta = \pi/2$.
The  left panel corresponds to  $v = 0.8$, the central panel to  $v = 0.9$ and the right panel corresponds to $v=1$.

For $v=0.8$ (left panel), the  most unstable modes are those corresponding to  small angles $\theta$, \ie are
those modes with $\bf k$  almost collinear with the
velocity of the jet.  In this regime these modes  dominate the dynamics.
With increasing velocity modes with larger values of the angle become relevant.
Indeed for  a value of the velocity $v = 0.9$, corresponding to the central panel in Fig.~\ref{obliquefig1}, the imaginary part
of the modes with $0<\theta < \pi/4$ is the largest and the corresponding modes dominant. In the ultrarelativistic case $v \to1$,    right panel in Fig.~\ref{obliquefig1},
the dominant modes are  those  with $\theta \sim \pi/4$. Modes with $\bf k$ almost collinear with $\bf v$, or more precisely all the modes at angles  $\theta \lesssim \pi/8$,
are suppressed. Notice that
the modes with very large values of $\theta \sim \pi/2$ are not the dominant one.
This means that modes with $\bf k$ orthogonal to $\bf v$  are not the most important:
more important are ``oblique" modes with $\theta \sim \pi/4$.

In agreement with the analytical results of the previous Section, in  the limit  $v \to 1$,  that is shown on the right panel of Fig.~\ref{obliquefig1},
the  mode with $\theta = 0$ has a vanishing imaginary part and is stable.

\section{Conclusions}
\label{conclusion-sec}

It is well-known that hydrodynamics predicts that any object moving in a fluid at a velocity
higher than the  speed of sound creates shock waves and a Mach cone structure. The conclusion
of our study is that  a neutral stream of colored particles moving
at a velocity higher than the speed of sound in a quark-gluon plasma in the conformal limit
also causes gauge field instabilities.  It is curious to note
that it is  the speed of sound the threshold value for the appearance of such instabilities. Qualitatively one can 
understand why this happens considering that if an external
perturbation to the plasma  moves slower than the speed of sound, the system can respond
by locally rearranging the values of density, energy density, pressure and plasma velocity, and then
hydrodynamical fluctuations of these quantities may  counteract the slow external perturbation. 
This is not the case
when the perturbation propagates faster than the larger  velocity of propagation of  hydrodynamical fluctuations, \ie larger
than the speed of sound.

Even if we have obtained our results by solving  the chromohydrodynamical equation of the system
in the linear response approximation 
let us stress that the same  results are  easily translated for  electromagnetic relativistic fluids. 
To the best of our knowledge, the speed of sound bound that we found has not 
been discussed in the existing literature.

We believe that the mechanism proposed in this article provides
one possible effective hydrodynamical description of the jet
quenching phenomena, valid at macroscopic scales. Indeed we
have shown that the energy and momenta stored in the total system
(composed by the plasma and the jet) is effectively converted into
(growing) energy and momenta stored in the gauge fields, which are
initially absent.  The jet looses energy in its dynamical evolution.

However, our analysis of this mechanism has to be considered as preliminary and therefore we cannot indicate
 consequences  for existing data of jet quenching at RHIC.
Indeed,  we have neglected  several aspects of the system produced in heavy-ion collisions  that complicate
the treatment of the jet quenching phenomenon,  such as the expansion of the plasma, 
or the transition to an hadronic phase, that  takes
place when the plasma becomes sufficiently dilute. One of the relevant points here is
whether the system has enough time to generate the growth of the gauge fields before  hadronization begins. As found in our numerical study,
the maximal value of the growth rate, here characterized by ${\Gamma}_{\rm max}$, runs between  ${\Gamma}_{\rm max} \sim (0.08-0.15) \,\omega_t$. Therefore, the plasma instabilities fully develop on time scales of the order
$t \sim (6.7- 12.5)/\omega_t$. To get an upper bound of that time scale, we evaluate the plasma frequency
in a weakly coupling scenario at $T \sim 350$ MeV, finding  $t \sim 1 - 2$ fm/c.  

Another aspect that we have neglected is the existence of  dissipative terms in the hydrodynamical equations that at a certain stage of the plasma evolution might be able to damp the hydrodynamical fluctuations.

In any case,   assuming that  the generated gauge fields have sufficiently time to grow, what is their fate? How do they hadronize? The late stage evolution of the chromohydrodynamical instabilities may depend on a different number of facts.
 One should discuss what is the saturation mechanism of the instabilities, which could be drastically affected by dissipative color damping phenomena. 
In any case, naively speaking, we would expect that the existence
of some gauge field modes at  soft values of the momenta less than $k_{\rm max}$, and in a given direction
in momenta space, should be translated at RHIC into an enhanced production of hadrons with momenta
 $< k_{\rm max}$, versus the case without gauge field instabilities. Moreover, if one takes into account  the effect of the growth of the gauge fields  in the numerical simulations of the conical flow then  this might  lead to an enhancement of  the number of soft hadrons  produced by the jet 
 with respect to the case     where the effect of the growth of the gauge fields has not been considered \cite{Chaudhuri:2005vc}.

\begin{acknowledgments}
We kindly thank Stanis\l aw Mr\' owczy\' nski and Carlos A. Salgado for
the critical reading of the manuscript and their useful remarks.
This work has been supported by the Ministerio de Educaci\'on y Ciencia (MEC) under grant
AYA 2005-08013-C03-02.

\end{acknowledgments}



\begin{thebibliography}{99}

\bibitem{Adams:2005dq}
  J.~Adams {\it et al.}  [STAR Collaboration],
  Nucl.\ Phys.\  A {\bf 757}, 102 (2005);
  B.~B.~Back {\it et al.},
  Nucl.\ Phys.\  A {\bf 757}, 28 (2005);
  I.~Arsene {\it et al.}  [BRAHMS Collaboration],
  Nucl.\ Phys.\  A {\bf 757}, 1 (2005);
%
  K.~Adcox {\it et al.}  [PHENIX Collaboration],
  Nucl.\ Phys.\  A {\bf 757}, 184 (2005)

\bibitem{Kolb:2003dz}
  P.~F.~Kolb and U.~W.~Heinz,
  ``Hydrodynamic description of ultrarelativistic heavy-ion collisions,''
  arXiv:nucl-th/0305084.


\bibitem{Kovner:2003zj}
  A.~Kovner and U.~A.~Wiedemann,
  arXiv:hep-ph/0304151;
  M.~Gyulassy, I.~Vitev, X.~N.~Wang and B.~W.~Zhang,
  arXiv:nucl-th/0302077;
  P.~Jacobs and X.~N.~Wang,
  Prog.\ Part.\ Nucl.\ Phys.\  {\bf 54}, 443 (2005)
  [arXiv:hep-ph/0405125].
\bibitem{Eskola:2004cr}
  K.~J.~Eskola, H.~Honkanen, C.~A.~Salgado and U.~A.~Wiedemann,
  Nucl.\ Phys.\  A {\bf 747}, 511 (2005)
  [arXiv:hep-ph/0406319].
  A.~Dainese, C.~Loizides and G.~Paic,
  Eur.\ Phys.\ J.\  C {\bf 38}, 461 (2005)
  [arXiv:hep-ph/0406201].

\bibitem{Liu:2006ug}
  H.~Liu, K.~Rajagopal and U.~A.~Wiedemann,
  Phys.\ Rev.\ Lett.\  {\bf 97}, 182301 (2006)
  [arXiv:hep-ph/0605178].
\bibitem{Chaudhuri:2005vc}
  A.~K.~Chaudhuri and U.~Heinz,
  Phys.\ Rev.\ Lett.\  {\bf 97}, 062301 (2006)
  [arXiv:nucl-th/0503028].

\bibitem{Stoecker:2004qu}
  H.~Stoecker,
  Nucl.\ Phys.\  A {\bf 750}, 121 (2005)
  [arXiv:nucl-th/0406018].

\bibitem{Casalderrey-Solana:2004qm}
  J.~Casalderrey-Solana, E.~V.~Shuryak and D.~Teaney,
  J.\ Phys.\ Conf.\ Ser.\  {\bf 27}, 22 (2005)
  [Nucl.\ Phys.\  A {\bf 774}, 577 (2006)]
  [arXiv:hep-ph/0411315];
  J.~Casalderrey-Solana, E.~V.~Shuryak and D.~Teaney,
  arXiv:hep-ph/0602183.
\bibitem{Adams:2005ph}
  J.~Adams {\it et al.}  [STAR Collaboration],
  Phys.\ Rev.\ Lett.\  {\bf 95}, 152301 (2005)
  [arXiv:nucl-ex/0501016]; 
  J.~G.~Ulery  [STAR Collaboration],
  Nucl.\ Phys.\  A {\bf 783}, 511 (2007)
  [arXiv:nucl-ex/0609047].

\bibitem{Adler:2005ee}
  S.~S.~Adler {\it et al.}  [PHENIX Collaboration],
  Phys.\ Rev.\ Lett.\  {\bf 97}, 052301 (2006)
  [arXiv:nucl-ex/0507004].

\bibitem{Polosa:2006hb}
  A.~D.~Polosa and C.~A.~Salgado,
  Phys.\ Rev.\  C {\bf 75}, 041901 (2007)
  [arXiv:hep-ph/0607295].

\bibitem{Honda}
M.~Honda, Phys.\ Rev. {\bf E 69}, 016401 (2004).

\bibitem{Bret}
A.~Bret and C.~Deutsch, Phys. \ Plasmas {\bf 13}, 042106 (2006).


\bibitem{exp-inst}
M. Tatarakis et al,  Phys.\ Rev.\ Lett.\  {\bf 90}, 175001 (2003).






\bibitem{Mrowczynski:1994xv}
  S.~Mrowczynski,
  Phys.\ Rev.\  C {\bf 49}, 2191 (1994).

\bibitem{Arnold:2004ti}
  P.~Arnold, J.~Lenaghan, G.~D.~Moore and L.~G.~Yaffe,
  Phys.\ Rev.\ Lett.\  {\bf 94}, 072302 (2005)
  [arXiv:nucl-th/0409068].


\bibitem{Mrowczynski:1988dz}
  S.~Mrowczynski,
  Phys.\ Lett.\  B {\bf 214}, 587 (1988).




\bibitem{Randrup:2003cw}
  J.~Randrup and S.~Mrowczynski,
  Phys.\ Rev.\  C {\bf 68}, 034909 (2003)
  [arXiv:nucl-th/0303021].

\bibitem{Romatschke:2003ms}
  P.~Romatschke and M.~Strickland,
  Phys.\ Rev.\  D {\bf 68}, 036004 (2003)
  [arXiv:hep-ph/0304092].

\bibitem{Arnold:2003rq}
  P.~Arnold, J.~Lenaghan and G.~D.~Moore,
  JHEP {\bf 0308}, 002 (2003)
  [arXiv:hep-ph/0307325].

\bibitem{Rebhan:2004ur}
  A.~Rebhan, P.~Romatschke and M.~Strickland,
  Phys.\ Rev.\ Lett.\  {\bf 94}, 102303 (2005)
  [arXiv:hep-ph/0412016].

\bibitem{Dumitru:2005gp}
  A.~Dumitru and Y.~Nara,
  Phys.\ Lett.\  B {\bf 621}, 89 (2005)
  [arXiv:hep-ph/0503121].

\bibitem{Arnold:2005vb}
  P.~Arnold, G.~D.~Moore and L.~G.~Yaffe,
  Phys.\ Rev.\  D {\bf 72}, 054003 (2005)
  [arXiv:hep-ph/0505212].

\bibitem{Rebhan:2005re}
  A.~Rebhan, P.~Romatschke and M.~Strickland,
  JHEP {\bf 0509}, 041 (2005)
  [arXiv:hep-ph/0505261].

\bibitem{Arnold:2005ef}
  P.~Arnold and G.~D.~Moore,
  Phys.\ Rev.\  D {\bf 73}, 025006 (2006)
  [arXiv:hep-ph/0509206].

\bibitem{Arnold:2005qs}
  P.~Arnold and G.~D.~Moore,
  Phys.\ Rev.\  D {\bf 73}, 025013 (2006)
  [arXiv:hep-ph/0509226].

\bibitem{Dumitru:2006pz}
  A.~Dumitru, Y.~Nara and M.~Strickland,
  Phys.\ Rev.\  D {\bf 75}, 025016 (2007)
  [arXiv:hep-ph/0604149].

\bibitem{Romatschke:2005pm}
  P.~Romatschke and R.~Venugopalan,
  Phys.\ Rev.\ Lett.\  {\bf 96}, 062302 (2006)
  [arXiv:hep-ph/0510121].

\bibitem{Romatschke:2005ag}
  P.~Romatschke and R.~Venugopalan,
  Eur.\ Phys.\ J.\  A {\bf 29}, 71 (2006)
  [arXiv:hep-ph/0510292].

\bibitem{Bodeker:2007fw}
  D.~Bodeker and K.~Rummukainen,
  arXiv:0705.0180 [hep-ph].


\bibitem{Mrowczynski:2006ad}
  S.~Mrowczynski,
  Proceedings of `Critical Point and Onset
of Deconfinement (CPOD2006)', July 3-6, 2006
Florence, Italy, PoS(CPOD2006)042 [arXiv:hep-ph/0611067].
  

\bibitem{Strickland:2007fm}
  M.~Strickland,
  ``Thermalization and the chromo-Weibel instability,''
  arXiv:hep-ph/0701238.


\bibitem{Manuel:2006hg}
  C.~Manuel and S.~Mrowczynski,
  Phys.\ Rev.\  D {\bf 74}, 105003 (2006)
  [arXiv:hep-ph/0606276].



\bibitem{Pavlenko:1991ih}
  O.~P.~Pavlenko,
  Sov.\ J.\ Nucl.\ Phys.\  {\bf 55}, 1243 (1992)
  [Yad.\ Fiz.\  {\bf 55}, 2239 (1992)].



\bibitem{Ruppert:2005uz}
  J.~Ruppert and B.~Muller,
  Phys.\ Lett.\  B {\bf 618}, 123 (2005)
  [arXiv:hep-ph/0503158].

\bibitem{Chakraborty:2006md}
  P.~Chakraborty, M.~G.~Mustafa and M.~H.~Thoma,
  Phys.\ Rev.\  D {\bf 74}, 094002 (2006)
  [arXiv:hep-ph/0606316].





\bibitem{Manuel:2003zr}
C.~Manuel and St.~Mr\'owczy\'nski,
Phys.\ Rev.\ D {\bf 68}, 094010 (2003).


\bibitem{Manuel:2004gk}
C.~Manuel and St.~Mr\'owczy\'nski,
Phys.\ Rev.\ D {\bf 70}, 094019 (2004).

\bibitem{foot1}
Notice that in principle we should have defined the quantity $(c_s^a)^2 = \delta p_a /\delta \epsilon_a$, however in the conformal limit $c_s^a=c_s$. 

\bibitem{Lebellac}
M. Le Bellac, ``Thermal Field Theory", (University Press, Cambridge, 1991).





\bibitem{Jackiw:1991ck}
  R.~Jackiw, D.~Kabat and M.~Ortiz,
  Phys.\ Lett.\  B {\bf 277}, 148 (1992)
  [arXiv:hep-th/9112020].

\end{thebibliography}
\end{document}